\documentclass[12pt]{article}%
\usepackage{amsfonts}
\usepackage{graphicx}
\usepackage{amsmath}
\usepackage{amssymb}%
\pdfoutput=1
\makeatletter

\@addtoreset{equation}{section}
\makeatother
\newcommand{\beq}{\begin{equation}}
\newcommand{\eeq}{\end{equation}}
\renewcommand{\a}{\alpha}
\renewcommand{\b}{{{\beta}}}

\newcommand{\be}{\begin{eqnarray}}
\newcommand{\ee}{\end{eqnarray}}

\begin{document}
\baselineskip=18pt
\baselineskip 0.7cm

\begin{titlepage}
\setcounter{page}{0}
\renewcommand{\thefootnote}{\fnsymbol{footnote}}
\begin{flushright}
\end{flushright}
\begin{center}
{\Large \bf
Chern-Simons Theory on Supermanifolds
}

\vspace{1cm}
{\large
Pietro Antonio Grassi$^{~a,b,}$\footnote{pietro.grassi@uniupo.it}
and 
Carlo Maccaferri$^{~c,b}$\footnote{maccafer@gmail.com}

\medskip
}
\vskip 0.5cm
{
\small
\centerline{$^{(a)}$ \it Dipartimento di Scienze e Innovazione Tecnologica,}
\centerline{\it Universit\`a del Piemonte Orientale,}
\centerline{\it viale T. Michel, 11, 15121 Alessandria, Italy.}
\medskip
\centerline{$^{(b)}$ \it INFN, Sezione di Torino,} 
\centerline{\it via P. Giuria 1, 10125 Torino.} 
\medskip
\centerline{$^{(c)}$ \it Dipartimento di Fisica, Universit\`a di Torino,}
\centerline{\it via P. Giuria , 1, 10125 Torino, Italy.}
}
\vskip  .3cm
\medskip
\end{center}
\smallskip
\centerline{{\bf Abstract}}
\medskip
\noindent
{
We consider quantum field theories  on supermanifolds using integral forms. 
The latter are used to define a geometric theory of integration  
and  they are essential for a consistent action principle. 
The construction relies on Picture Changing Operators, analogous to the one 
introduced in String Theory. 
As an application, we construct a geometric action principle for N=1 D=3 super-Chern-Simons theory.   
}
\vskip  .5cm
\centerline{\it Dedicated to the memory of Raymond Stora.}
\vskip  1cm

\noindent
{\today}
\end{titlepage}
\setcounter{page}{1}


\vfill
\eject
\newpage\setcounter{footnote}{0} \newpage\setcounter{footnote}{0}

\section{Introduction}

One of the main differences between the geometry of supermanifolds and that of conventional manifolds 
is the distinction between differential forms and integral forms \cite{Catenacci:2010cs,Castellani:2014goa}. 
The latter are essential to provide a geometric integration theory for supermanifolds. 
Since the differentials $d\theta$'s, associated to anticommuting coordinates $\theta$'s,
are commuting variables there is no natural integrable top-differential form. Then one introduces
distribution-like anti-commuting quantities, such as for example $\delta(d\theta)$, that can provide a suitable integral top-form and for which the usual Cartan calculus can be extended (see here fore a non-exhaustive reference list \cite{Voronov2,integ,Witten:2012bg}). Therefore, the complex of the differential forms together with the complex of the 
integral forms (those which can be integrated) are the highest and the lowest line of the interesting double complex of the 
{\it pseudo-forms}. 

The complex, whose elements are denoted by $\Omega^{(p|q)}$,  
 is filtered by two integer numbers: $p$, which represents the usual {\it form degree} (which can also be negative) and $q$, {\it the picture number}, which counts the number of delta functions and it ranges between $0$ and $m$,  with $m$ the fermionic dimension of manifold.  It is customary to denote by {\it superforms} those with vanishing picture $\Omega^{(p|0)}$ with no bound on the form degree; while the {\it integral forms} are those in $\Omega^{(p|m)}$. An integral form 
of top degree can be integrated on a supermanifold and it produces a number like a usual differential form does on a manifold. 
The differential $d$, suitably extended to the entire complex, increases the form degree without touching the picture number. The latter can be modified  
 by increasing and lowering the number of delta functions, and for that  one needs new operators  known as 
 {\it picture changing operators} PCO's originally introduced  in RNS string theory \cite{Friedan:1985ge}. There,  the role of the supermanifold is played by the worldsheet super-Riemann surface, or more precisely by the associated super-moduli space and super-conformal Killing group, 
 as discussed in \cite{Witten:2012bg} and 
 integral forms are essential to define the amplitudes to all orders of the genus expansion. In higher dimensional spacetime theory, but based on worldsheet two-dimensional models, they were introduced in \cite{Berkovits:2004px} and further discussed in \cite{Grassi:2004tv}.

In the present paper, we discuss the role of PCO in the context of spacetime QFT and the relation between different superspace formalisms. All of them are related by a choice of suitable PCO with different properties, but belonging to 
the same cohomology class. As a playground, we choose 3D, N=1 super-Chern-Simons theory. 
 
The conventional bosonic Chern-Simons theory is described by the geometrical action
\begin{eqnarray}
\label{introA}
S_{CS} = \int_{\cal M} {\rm Tr}\Big( A^{(1)} \wedge d A^{(1)} + \frac{2}{3} A^{(1)}\wedge A^{(1)}\wedge A^{(1)} \Big)\,,
\end{eqnarray}
where $A^{(1)}$ is the 1-form gauge connection with values in the adjoint representation of the 
gauge group ${\cal G}$, 
the trace is taken over the same representation and the integral integrates a $3$-form Lagrangian 
over a three dimensional manifold ${\cal M}$. As is well 
known, it provides a meaningful integral, independent of the parametrization of ${\cal M}$ and of its metric. The 
$3$-form Lagrangian is closed by construction and its gauge variation is exact. 

For the corresponding {\it super} Chern-Simons action on a supermanifold ${\cal M}^{(3|2)}$
one needs a $(3|2)$-integral form that, however, cannot be built only by connections as $A^{(1|0)}$. 
The latter are differential $1$-superform with zero picture 
 (as been explained in \cite{Catenacci:2010cs,Castellani:2014goa}), leading to a $(3|0)$ superform Lagragian as 
 (\ref{introA}) that cannot be integrated. Nonetheless, it can be converted to a $(3|2)$-integral form by multiplying it by a PCO 
 belonging to $\Omega^{(0|2)}$ for example 
 \begin{eqnarray}
\label{introB}
\mathbb{Y}^{(0|2)}_{new} = V^a \wedge V^b (\gamma_{ab})^{\alpha\beta} \iota_\alpha \iota_\beta \delta^2(d\theta)\,,
\end{eqnarray}
where $V^a = dx^a + \theta^\alpha \gamma^a_{\alpha\beta} d\theta^\beta$ is the super-line element. $\gamma^a_{\alpha\beta}, 
\gamma^{ab}_{\alpha\beta}$ are the Dirac gamma matrices and $\iota_\alpha$ is the usual contraction operators along 
the odd vector $D_\a = \partial_\alpha - (\theta \gamma^a)_\beta \partial_a$. The operator 
$\mathbb{Y}^{(0|2)}_{new}$ is closed, 
supersymmetric  and not exact, namely it belongs to $H^{(0|2)}$. 

Consequently the super Chern-Simons action reads 
\begin{eqnarray}
\label{introC}
S_{SCS} = \int_{\cal SM} \mathbb{Y}^{(0|2)}_{new} \wedge 
{\rm Tr}\Big( A^{(1|0)} \wedge d A^{(1|0)}  + \frac{2}{3} A^{(1|0)}\wedge A^{(1|0)}\wedge A^{(1|0)}\Big)\,,
\end{eqnarray}
where the integration is extended to the entire supermanifold ${\cal SM}$. As can be checked, the result is 
gauge invariant, supersymmetric and leads to the well-known super Chern-Simons action in superspace. An obvious question is whether one can change the PCO  $\mathbb{Y}^{(0|2)}_{new}$ without changing the 
action. Since $ \mathbb{Y}^{(0|2)}_{new}$ belongs to a cohomology class, it implies a choice of a representative inside 
of the same class. This means that the invariance of the action w.r.t. to a change of $ \mathbb{Y}^{(0|2)}_{new}$ 
is achievable only if the $(3|0)$ Lagrangian is closed. That request, for a $(3|0)$ superform in the supermanifold ${\cal M}^{(3|2)}$, is non-trivial and indeed the action given in (\ref{introC}) has to be modified accordingly. It is easy to show that there is a missing term in the action and the closure implies the usual conventional constraints. Then, after that modification, we can change the PCO for getting new forms of the action with the same physical content, but displaying different properties. 
 
In the present context, we provide a new geometrical perspective on QFT's superspace and on supermanifolds. We are able to prove that the {\it Rheonomic} action (see \cite{Castellani}) formulation of $N=1$ $D=3$ super Chern-Simons theory with rigid supersymmetry (the local supersymmetric case will be discussed separately) can be considered  a ``mother" theory which has built-in all possible superspace realizations for that theory. In particular we show that using a given PCO the action reduces to the usual action in terms of component fields and by another choice we get the superspace action written in terms of superfields. However, only for the choice (\ref{introB}) we are able to derive the conventional constraint by varying the action and without resorting to the rheonomic parametrization.
 
 The paper is organised as follows: Sec. 2 deals with background material, the definition of integral forms and integration on 
 supermanifolds. In Sec. 3, we introduce PCO's for spacetime quantum field theory. In Sec. 4, we discuss the action of super-Chern-Simons theory in 3d. The relation between different types of PCO's and actions are given in Sec. 5.  

Integral forms, integration on supermanifolds, the role of picture changing operators in QFT
 and applications to gauge theories was one of the last discussions with Raymond Stora during the 
 last extended period spent by one of the authors at CERN, for that reason this note is dedicated to him. 
 
\section{Background Material}
\subsection{$3d, N=1$}

We recall that in 3d N=1, the supermanifold $\mathcal{SM}^{(3|2)}$ (homeomorphic
to $\mathbb{R}^{3|2}$) is described locally by the coordinates $(x^{a}%
,\theta^{\a})$, and in terms of these coordinates, we have the following two
differential operators
\begin{equation}
D_{\a}=\frac{\partial}{\partial \theta^{\a}}-(\gamma^{a}\theta)_{\a}\partial_{a}%
\,,~~~~~~Q_{\a}=\frac{\partial}{\partial \theta^{\a}} + (\gamma^{a}\theta)_{\a}\partial_{a}\,,~~~~~~
\label{susy3dA}%
\end{equation}
known as superderivative and supersymmetry generator, respectively. They have the
properties\begin{equation}
\{D_{\a},D_{\b} \}=-2 \gamma^a_{\a\b} \partial_a\,, ~~~~~~
\{Q_{\a},Q_{\b} \}= 2 \gamma^a_{\a\b} \partial_a\,, ~~~~~~
\{D_{\a},Q_{\b} \}=0\,,
\label{susy3dB}
\end{equation}
In 3d, with $\eta_{ab} = (-,+,+)$, 
we use real and symmetric Dirac matrices $\gamma^a_{\a\b}$ defined as 
 \begin{eqnarray}
\label{dicA}
&&\gamma^0_{\a\b} = (C \Gamma^0) = - {\mathbf 1}\,, ~~~
\gamma^1_{\a\b} = (C \Gamma^1) = \sigma^3\,,  \nonumber \\
&&\gamma^2_{\a\b} = (C \Gamma^2) = - \sigma^1\,, ~~
C_{\a\b} = i \sigma^2 = \epsilon_{\a\b}\,. 
\end{eqnarray}
Numerically, we have $\hat\gamma_a^{\a\b} = \gamma^a_{\a\b}$ and 
$\hat\gamma_a^{\a\b} = \eta_{ab} (C \gamma^b C)^{\a\b} = C^{\a\gamma} \gamma_{a, \gamma\delta} C^{\delta\beta}$. 
The conjugation matrix is
$\epsilon^{\a\b}$ and a bi-spinor is decomposed as follows $R_{\a\b} = R \epsilon_{\a\b}  + R_a \gamma^a_{\a\b}$ where
$R = - \frac12 \epsilon^{\a\b} R_{\a\b}$ and $R_a = {\rm tr}(\gamma_a R)$ are
a scalar and a vector, respectively.
In addition, it is easy to show that
$\gamma^{ab}_{\a\b} \equiv \frac12 [\gamma^a, \gamma^b] = \epsilon^{abc} \gamma_{c \a\b}$.

For computing the differential of $\Phi^{(0|0)}$, we can use the basis of $(1|0)$-forms defined as follows
\begin{equation}
d \Phi^{(0|0)} = dx^{a} \partial_{a} \Phi^{(0|0)} + d\theta^{\alpha}%
\partial_{\alpha}\Phi^{(0|0)} =
\end{equation}
\[
=\Big(dx^{a} + \theta\gamma^{a} d\theta\Big) \partial_{a} \Phi^{(0|0)}
+ d\theta^{\alpha}D_{\alpha}\Phi^{(0|0)} \equiv V^{a} \partial_{a}
\Phi^{(0|0)} + \psi^{\alpha}D_{\alpha}\Phi^{(0|0)}\,,
\]
where $V^a = dx^a + \theta \gamma^a d\theta$ and $\psi^\a = d \theta^\a$ which satisfy the Maurer-Cartan 
equations 
\begin{eqnarray}
\label{}
d V^a = \psi \gamma^a \psi\,, ~~~~~~ d \psi^\a = 0\,. 
\end{eqnarray}

Given a $(0|0)$-form $\Phi^{(0|0)}$, we can compute its supersymmetry variation (viewed as a super translation)
as a Lie derivative $\mathcal{L}_{\epsilon}$ with $\epsilon=\epsilon
^{\a}Q_{\a}+\epsilon^{a}\partial_{a}$ ($\epsilon^{a}=\epsilon^\alpha\gamma^a_{\alpha\beta}\epsilon^\beta$ are the infinitesimal
parameters of the translations and $\epsilon^{\a}$ are the supersymmetry
parameters) and we have
\begin{equation}
\delta_{\epsilon}\Phi^{(0|0)}=\mathcal{L}_{\epsilon}\Phi^{(0|0)}%
=\iota_{\epsilon}d\Phi^{(0|0)}=\iota_{\epsilon}\Big(dx^{a}\partial_{a}%
\Phi^{(0|0)}+d\theta^{\a}\partial_{\a}\Phi^{(0|0)}\Big)=
\end{equation}%
\[
=(\epsilon^{a}+\epsilon\gamma^{a}\theta)\partial_{a}\Phi
^{(0|0)}+\epsilon^{\a}\partial_{\a}\Phi^{(0|0)}=\epsilon^{a}\partial_{a}%
\Phi^{(0|0)}+\epsilon^{\a}Q_{\a}\Phi^{(0|0)}\,,%
\]
In the same way, acting on $(p|q)$ forms, where $p$ is the form degree and $q$
is the picture number, we use the usual Cartan formula $\mathcal{L}_{\epsilon
}=\iota_{\epsilon}d+d\iota_{\epsilon}$. It follows easily that 
$\delta_{\epsilon}V^{a} = \delta_{\epsilon}V^{\alpha}=0$ and  $\delta_{\epsilon}d \Phi^{(0|0)} = d
\delta_{\epsilon}\Phi^{(0|0)}$.

The top form is represented by the expression
\begin{equation}
\label{top3d}\omega^{(3|2)} = \epsilon_{abc} V^{a}\wedge V^{b} \wedge V^{c} \wedge\epsilon_{\alpha\beta} \delta(\psi^{\alpha}) \wedge
\delta(\psi^{\beta})\,,
\end{equation}
which has the properties
\begin{equation}
\label{top3dB}d \omega^{(3|2)} = 0\,, ~~~~~ \mathcal{L}_{\epsilon}
\omega^{(3|2)} =0\,.
\end{equation}

It is important to point out the transformation properties of $\omega^{(3|2)}$ under a Lorentz transformation of $SO(2,1)$.
Considering $V^a$, which transforms in the vector representation of $SO(2,1)$, the combination $\epsilon_{abc} V^{a}\wedge V^{b} \wedge V^{c}$ is clearly invariant. On the other hand, $d\theta^\a$ transform under the spinorial
representation of $SO(2,1)$, say $\Lambda_\a^{~\b} = (\gamma^{ab})_\a^{~\b}  \Lambda_{ab}$ with $\Lambda_{ab} \in so(2,1)$, and thus
an expression like $\delta(d\theta^\a)$ is not covariant. Nonetheless, the combination $\epsilon^{\a\b} \delta(d\theta^\a) \delta(d\theta^\b) = 2 \delta(d\theta^1) \delta(d\theta^2)$ is invariant using formal mathematical properties of 
distributions, for instance $d\theta \delta(d\theta)$ and $d\theta \delta'(d\theta) = - \delta(d\theta)$. We  recall that 
$\delta(\psi^\a) \wedge \delta(\psi^\b) = - \delta(\psi^\b) \wedge \delta(\psi^\a)$. 
In addition, $\omega^{(3|2)}$ has a bigger symmetry group: we can transform the variables ($V^\a, d\theta^\a)$ under an element of the supergroup $SL(3|2)$. The form $\omega^{(3|2)}$ is a representative of the Berezinian bundle, the
equivalent for supermanifolds of the canonical bundle on bosonic manifolds.

\subsection{Integral Forms}

Consider the generalized form multiplication as 
\begin{equation}\label{inNA}
\wedge: \Omega^{(p|r)} ({\cal SM}) \times \Omega^{(q|s)} ({\cal SM}) \longrightarrow \Omega^{(p+q|r+s)} ({\cal SM})\,. 
\end{equation}
where $0\leq p,q \leq n$ and $0\leq r,s \leq m$ with $(n|m)$ are the bosonic and fermonic dimensions of the 
supermanifold ${\cal SM}$. Due to the anticommuting properties of the delta forms this product is by definition equal to zero if the forms to be multiplied contain delta forms localized in the same variables $d\theta$. 

Given the space of pseudo forms 
$\Omega^{(p|r)}$, a $(p|r)$-form $\omega$ formally reads 
\begin{equation}\label{inNBA}
\omega = \sum_{l,h,r} \omega_{[a_1 \dots a_l] (\a_{1} \dots \a_{h}) [\b_{1} \dots \b_{r}]}  dx^{a_1} \dots dx^{a_l} 
d\theta^{\a_1} \dots d\theta^{\a_h} \delta^{^{g(\beta_1)}}(d\theta^{\b_1})   \dots_\wedge \delta^{^{g(\beta_r)}}(d\theta^{\b_r})\,, 
\end{equation}
where $g(t)$ denotes the differentiation degree of the Dirac delta function corresponding to the 1-form 
$d\theta^t$.\footnote{It is an easy exercise to rewrite $\omega$ in terms of the susy invariant superforms $V^a, \psi^\a$.} If $g(t)=0$ it means that the Dirac delta function has no derivative. The 
three indices $l, h$ and $r$ satisfy the relation 
\begin{equation}\label{inNBB}
l + h - \sum_{k=1}^r g(\b_k) = p\,, ~~~~~~\a_l \neq \{\b_1, \dots, \b_r\} ~~~ \forall l=1,\dots,h\,,
\end{equation}
where the last equation means that each $\alpha_l$ in the above summation should be different from any $\beta_k$, otherwise the 
degree of the differentiation of the Dirac delta function can be reduced and the corresponding 1-form $d\theta^{\a_k}$ is removed from the basis. 
 The components  $\omega_{[i_1 \dots i_l] (\a_{1} \dots \a_{m}) [\b_{1} \dots \b_{r}]}$  
of $\omega$ are superfields. 

In fig. 1, we display the complete complex of pseudo-forms. 
We notice that the first line and the last line are bounded from below and from above, respectively. This is due to the fact that in the first line, being absent any delta functions, 
the form number cannot be negative, and in the last line, having saturated the number of delta functions we cannot admit any power of $d\theta$ (because of the 
distributional law $d\theta \delta(d\theta) =0$). 

Before discussing the Chern-Simons action, we analyze the dimension of each space $\Omega^{(p|r)}$. 
The dimension of $\Omega^{(p|0)}$ is given by the power of the $dx$ 1-forms and by the power of the $d\theta$ 1-form
\begin{equation}\label{in NBDA}
dx^{a_1} \dots dx^{a_l} d\theta^{\a_1} \dots d\theta^{\a_h}\,, 
\end{equation}
where we have decomposed the form degree $p$ into $l +h$ 
where the degree $l$ is carried by $dx$ and the degree $h$ is carried by $d\theta$. For that decomposition, we have $n(n-1) \dots (n-l+1)/l!$ components coming from 
$dx^{a_1} \dots dx^{a_l}$ plus $(m+h-1)(m+h-2) \dots m/h!$ coming from $d\theta^{\a_1} \dots d\theta^{\a_h}$. In the same way, 
if we consider the integral forms $\Omega^{(n-p|m)}$ of the last line, we see that we can have powers of $dx$ and derivatives on the Dirac delta functions as
\begin{equation}\label{inNBE}
dx^{i_1} \dots dx^{i_l} \delta^{g(\a_1)}(d\theta^{\a_1}) \dots \delta^{g(\a_m)}(d\theta^{\a_m})\,, 
\end{equation}
where $g(t)$ is the order of the derivative on $\delta(t)$. The form degree is $l - \sum_{k=1}^m g(\a_k)$. 
 
 \def\de{\mathrm{d}}
\begin{figure}[t]
\begin{center}
\begin{tabular}{c@{\hskip -0mm}c@{\hskip 0mm}c}
\begin{tabular}{c@{\hskip 1.5mm}c@{\hskip 0.1mm}c}
&&\\[-0.34cm]
&\hspace{0.7cm}$0$&$\stackrel{\de}{\longrightarrow}$\\[0.06cm]
&\hspace{0.655cm}\tiny{$Z$}\hspace{0.3mm}\large{$\uparrow$}\phantom{\tiny{$Z$}}&\\[-0.05cm]
&\hspace{0.68cm}\vdots&\\[-0.05cm]
$\cdots$&$\Omega^{(-1|s)}$&$\stackrel{\de}{\longrightarrow}$\\[-0.15cm]
&\hspace{0.68cm}\vdots&\\[0.05cm]
&\hspace{0.655cm}\tiny{$Z$}\hspace{0.3mm}\large{$\uparrow$}\phantom{\tiny{$Z$}}&\\[0.05cm]
$\cdots$&$\Omega^{(-1|m)}$&$\stackrel{\de}{\longrightarrow}$\\[0.08cm]
\end{tabular}&
\begin{tabular}{|@{\hskip 1.5mm}c@{\hskip 1mm}c@{\hskip 2mm}c@{\hskip 1.5mm}c@{\hskip 1.5mm}c@{\hskip 1.5mm}c@{\hskip 1.5mm}c@{\hskip 1.5mm}|}
\hline
&&&&&&\\[-0.34cm]
$\Omega^{(0|0)}$&$\stackrel{\de}{\longrightarrow}$&$\cdots$&$\Omega^{(r|0)}$ &$\cdots$&$\stackrel{\de}{\longrightarrow}$&$\Omega^{(n|0)}$\\[0.06cm]
\tiny{$Z$}\hspace{0.3mm}\large{$\uparrow$}\large{$\downarrow$}\hspace{0.5mm}\tiny{$Y$}&&&\tiny{$Z$}\hspace{0.3mm}\large{$\uparrow$}\large{$\downarrow$}\hspace{0.5mm}\tiny{$Y$}&&&\tiny{$Z$}\hspace{0.3mm}\large{$\uparrow$}\large{$\downarrow$}\hspace{0.5mm}\tiny{$Y$}\\[-0.05cm]
\vdots&&&$\vdots$&&&$\vdots$\\[-0.05cm]
$\Omega^{(0|s)}$&$\stackrel{\de}{\longrightarrow}$&$\cdots$&$\Omega^{(r|s)}$ &$\cdots$&$\stackrel{\de}{\longrightarrow}$&$\Omega^{(n|s)}$\\[-0.15cm]
\vdots&&&$\vdots$&&&$\vdots$\\[0.05cm]
\tiny{$Z$}\hspace{0.3mm}\large{$\uparrow$}\large{$\downarrow$}\hspace{0.5mm}\tiny{$Y$}&&&\tiny{$Z$}\hspace{0.3mm}\large{$\uparrow$}\large{$\downarrow$}\hspace{0.5mm}\tiny{$Y$}&&&\tiny{$Z$}\hspace{0.3mm}\large{$\uparrow$}\large{$\downarrow$}\hspace{0.5mm}\tiny{$Y$}\\[0.05cm]
$\Omega^{(0|m)}$&$\stackrel{\de}{\longrightarrow}$&$\cdots$&$\Omega^{(r|m)}$&$\cdots$ &$\stackrel{\de}{\longrightarrow}$&$\Omega^{(n|m)}$\\[0.08cm]
\hline
\end{tabular}
&
\begin{tabular}{c@{\hskip 2mm}c@{\hskip 1.5mm}c}
&&\\[-0.34cm]
$\stackrel{\de}{\longrightarrow}$&$\Omega^{(n+1|0)}$&$\cdots$\\[0.06cm]
&\hspace{-1.15cm}\phantom{\tiny{$Y$}}\hspace{0.5mm}\large{$\downarrow$}\hspace{0.5mm}\tiny{$Y$}&\\[-0.05cm]
&\hspace{-1.15cm}\vdots&\\[-0.05cm]
$\stackrel{\de}{\longrightarrow}$&$\Omega^{(n+1|s)}$&$\cdots$\\[-0.15cm]
&\hspace{-1.15cm}\vdots&\\[0.05cm]
&\hspace{-1.15cm}\phantom{\tiny{$Y$}}\hspace{0.5mm}\large{$\downarrow$}\hspace{0.5mm}\tiny{$Y$}&\\[0.05cm]
$\stackrel{\de}{\longrightarrow}$&\hspace{-1.15cm}$0$&\\[0.08cm]
\end{tabular}
\end{tabular}
\end{center}\vskip -0.2cm\caption{\rm \scriptsize Structure of the supercomplex of forms on a supermanifold of dimension $(m|n)\,$. The form degree $r$ changes going from left to right while the picture degree $s$ changes going from up to down. The rectangle contains the subset of the supercomplex where the various pictures are isomorphic. In particular the de Rham cohomology is contained in square-box and each line is isomorphic to the other. }\label{scomplex}
\end{figure}

For example, for $n=3, m=2$ the superspace is ${\cal SM}^{(3|2)}$ and there are three complexes: $\Omega^{(p|0)}, \Omega^{(p|1)}$ and $\Omega^{(p|2)}$. 
The first one is bounded from below being $\Omega^{(0|0)}$ the lowest space generated by constant functions, the last one is bounded from above with $\Omega^{(3|2)}$ the highest 
space spanned by the top form and finally, the middle one is unbounded. In addition, the dimension of each space of the first and of the last one is 
finite, while for the middle one each $\Omega^{(p|1)}$ is infinite dimensional. 

Let us consider the space $\Omega^{(1|0)}$ spanned by ${dx^a, d\theta^\a}$ 
with dimensions $(3|2)$ (which means 3 bosonic generators -- instead of $dx^a$, one can use the supersymmetric variables
$V^a = dx^a + \theta \gamma^a d\theta$ -- and 2 fermionic generators $\psi^\a$). The space 
$\Omega^{(2|2)}$, spanned by 
$$\Big\{\epsilon_{a b c} dx^b dx^c \delta^2(d\theta), \epsilon_{a b c} dx^a dx^b dx^c \iota_\a \delta^2(d\theta)\Big\}\,,$$
where $\iota_\a \delta^2(d\theta)$ denote the derivative of $\delta^2(d\theta)$ with respect $d\theta^\a$.
It has dimensions $(3|2)$ and therefore there should be an isomorphism between the two spaces. The construction 
of that isomorphism, which is the generalization of the conventional Hodge dual to supermanifolds, has been provided in \cite{Castellani:2015ata}. 

Let us consider another example: the space $\Omega^{(2|0)}$ is spanned by 
$$\Big\{ \epsilon_{a b c} dx^{b}  dx^{c},  dx^a d\theta^\a,  d\theta^{(\a_1} d\theta^{\a_2)} \Big\}\,,$$ with dimension 
$(6|6)$. The dual space is $\Omega^{(1|2)}$ and it is 
spanned by 
$$\Big\{ dx^a \delta^2(d\theta), \epsilon_{a b c} dx^{b} dx^c \iota_\a \delta^2(d\theta), \epsilon_{a b c} dx^{a} dx^{b} dx^c \iota_{(\a_1} \iota_{\a_2)} \delta^2(d\theta)\Big\}\,,$$
which has again $(6|6)$ dimensions. 
The last example is the 
one-dimensional space $\Omega^{(0|0)}$ of $0$-forms and its dual $\Omega^{(3|2)}$, a one-dimensional space generated by $d^3x \delta^2(d\theta)$, the top form 
of the supermanifold ${\cal SM}^{(3|2)}$. 

Now, let consider the middle complex $\Omega^{(1|1)}$ spanned (in the sense of formal series) by the following 
psuedo-forms 
\begin{eqnarray}\label{midA}
\Omega^{(1|1)} &=& {\rm span} \Big\{(d\theta^\a)^{n+1} \delta^{(n)}(d\theta^\b), dx^{a} (d\theta^\a)^n \delta^{(n)}(d\theta^\b), \nonumber \\
&& \epsilon_{a b c }dx^{b} dx^{c} 
(d\theta^\a)^n \delta^{(n+1)}(d\theta^\b), \epsilon_{a b c } dx^a dx^{b} dx^{c}  (d\theta^\a)^n \delta^{(n+2)}(d\theta^\b)\Big\}_{n \geq 0}\,, 
\end{eqnarray}
where the number $n$ is not fixed and it must be a non-negative integer. Due to the bosonic 1-forms $dx^a$ and due to the fact that the index $\a$ must be different 
from $\b$ for a non-vanishing integral form (we recall that $d\theta^\a \delta^{(n)}(d\theta^\a) = - n \delta^{(n-1)}(d\theta^\a)$, and $\delta^{(0)}(d\theta^\a) = 
\delta(d\theta^\a)$), 
the number of generators (monomial forms) at a given $n$ is $(8|8)$, but the total 
number of monomial generators in $\Omega^{(1|1)}$ is infinite. The dual of $\Omega^{(1|1)}$ is itself, but the isomorphism is realised by 
an infinite matrix  whose entries are $(8|8) \times (8|8)$ supermatrices. 

In the same way, for a general supermanifold ${\cal M}^{(n|m)}$ 
any form belonging to the middle complex $\Omega^{(p|r)}$ with $0 < r < m$ is decomposed into 
an infinite number of components as in (\ref{midA}). 

In general, if $\omega$ is a poly-form in $\Omega^{\bullet}(\mathcal{M})$ this can be 
written as direct sum of $(p|q)$ pseudo forms
\begin{eqnarray}
\label{bimboA}
\omega = \sum_{p, q=0,1,2} \omega^{(p|q)}\,, 
\end{eqnarray}
and its
integral on the supermanifold is defined as follows: (in analogy with the
Berezin integral for bosonic forms):
\begin{equation}
\int_{\mathcal{SM}}\omega
\equiv\int_{\mathcal{M}}\epsilon^{i_{1}\dots i_{n}}\epsilon^{\beta_{1}%
\dots\beta_{m}}\omega_{\lbrack i_{1}\dots i_{n}][\beta_{1}\dots\beta_{m}%
]}(x,\theta)[d^{n}x\,d^{m}\theta]\,,\label{inLAA}%
\end{equation}
where the last integral over $\mathcal{M}$ is the usual Riemann-Lebesgue
integral over the coordinates $x^{i}$ (if it  exists) and the Berezin
integral over the coordinates $\theta^{\alpha}$. The superfields
$\omega_{\lbrack i_{1}\dots i_{n}][\beta_{1}\dots\beta_{m}]}(x,\theta)$ are
the components of the integral form and the symbol $[d^{n}x\,d^{m}\theta]$ denotes the 
integration variables.


\section{Picture Raising Operator}

In the present section, we discuss a class of PCO's  relevant to the study of differential forms in $\Omega^{(p|q)}$. 
 In particular we define a new operator  that increases 
the number of delta's (then, increases the picture number), the {\it Picture Raising Operator}.\footnote{We warn the reader the meaning of  raising and lowering is opposite to that used in string theory literature. In that case 
the picture is carried by the delta of the superghost $\delta(\gamma) = e^{-\phi}$ and it is conventionally taken to be negative, and indentified with the $\phi$ charge. 
} It acts vertically mapping superforms into integral forms.


To start with, 
given a constant commuting vector $v^\a$, consider the following object
\begin{equation}\label{PCOa}
Y_v = v \cdot \theta \, \delta(v \cdot d \theta)\,,
\end{equation}
 which has the properties
 \begin{equation}\label{PCOba}
d Y_v = 0\,, ~~~~~Y_v \neq d \eta^{(-1|1)}\,, ~~~~~  Y_{v + \delta v} = Y_v + d \Big(\delta v \cdot \theta \,  v \cdot \theta \delta'(v \cdot d\theta) \Big)\,,
\end{equation}
where $\eta^{(-1|1)}$ is a pseudo-form. Notice that $Y_v$ belongs to $H^{(0|1)}$ (which is the de-Rham 
cohomology class in $\Omega^{(0|1)}$) and by choosing two independent vectors $v_{(\a)}$, we have
\begin{equation}\label{PCOb}
\mathbb {Y}^{(0|2)} = \prod_{\a=1}^2 Y_{v_{(\a)}} =\theta^2 \delta^2(d\theta)\,,
\end{equation}
where $v^{\b}_{(\a)}$ is the $\b$-component of the vector $v_{(\alpha)}$. The result  is independent of $v^\a$. 
We can apply the PCO operator to a given integral form by taking the wedge product of forms. 
For example, given $\omega$ in $\Omega^{(p|0)}$ we have
\begin{equation}\label{PCOc}
\omega \longrightarrow \omega\wedge {\mathbb Y}^{(0|2)} =   {\mathbb Y}^{(0|2)} \wedge  \omega \in \Omega^{{(p|2)}}\,. 
\end{equation}
If $d \omega =0$ then $d (\omega \wedge {\mathbb Y}^{(0|2)}) =0$ (by applying the Leibniz rule), and if 
$\omega \neq d \eta$
then it follows that also $\omega \wedge {\mathbb Y}^{(0|2)} \neq d U$ where $U$ is an integral form of $\Omega^{(p-1|2)}$.
In \cite{Catenacci:2010cs}, it has been proved that ${\mathbb Y}^{(0|2)}$ is an element of the de Rham cohomology and
that they are also globally defined. So, given an element of the cohomogy $H_d^{(p|0)}$,
the new integral form $\omega \wedge {\mathbb Y}^{(0|2)}$ is an element of $H_d^{(p|2)}$.

Let us consider again the example of ${\cal M}^{(3|2)}$ and the 2-form $F^{(2|0)} = d A^{(1|0)} \in \Omega^{(2|0)}$
where $A^{(1|0)} = A_a V^a + A_\a \psi^\a \in \Omega^{(1|0)}$ is an abelian connection.
Then, we have 
\begin{equation}\label{PCOd}
F^{(2|0)} \longrightarrow \widetilde F^{(2|2)} = F^{(2|0)} \wedge {\mathbb Y}^{(0|2)}\,, 
\end{equation}
which satisfies the Bianchi identity $d \widetilde F^{(2|2)} =0$. 

Since the curvature $\widetilde {F}^{(2|2)} =F^{(2|0)} \wedge {\mathbb Y}$ can be also written as
$d A^{(1|0)} \wedge {\mathbb Y}^{(0|2)}$, using $d {\mathbb Y}^{(0|2)} =0$, we have
$$
 \widetilde F^{(2|2)} = d \left( A^{(1|0)} \wedge {\mathbb Y}^{(0|2)} \right)  = d \widetilde A^{(1|2)}\,,
$$
where ${\widetilde A}^{(1|2)}$ is the gauge connection at picture number 2.\footnote{Notice that besides the cases 
$A^{(1|0)}$ and $A^{(1|2)}$, we can also consider the case with one picture $A^{(1|1)}$, that would be the natural way 
to distribute the picture for CS theory. This shares similarities with open super string field theory in the $A_\infty$ formulation \cite{EKS} and it would be interesting to explore this further.} 
Notice that performing a gauge transformation on $A^{(1|0)}$, we have
$$\delta \widetilde{A}^{(1|2)} = d \left(\lambda^{(0|0)} \wedge {\mathbb Y}^{(0|2)}\right) \,, $$
and we can consider $\widetilde\lambda^{(0|2)} = \lambda^{(0|0)} \wedge {\mathbb Y}^{(0|2)}$ 
as the gauge parameter at picture number 2.

At the end, we have
\begin{equation}\label{PCOe}
F^{(2|0)} \wedge {\mathbb Y}^{(0|2)}
=
\Big(\partial_a A_b V^a V^b + \dots + (D_\a A_\b + \gamma^a_{\a\b} A_a) d\theta^\a d\theta^\b\Big) \wedge 
{\mathbb Y}^{(0|2)}
\end{equation}
$$
=( \partial_a A_b \theta^2)  \, V^a V^b \delta^2(d\theta) =
\partial_{[a} (A_{b]}(x,0) \theta^2)  \, V^a V^b \delta^2(d\theta) \,, 
$$
where $A_a(x,0)$ is the lowest component of the superfield $A_a$ appearing in the superconnection $A^{(1|0)}$.
This seems puzzling since we have ``killed" the complete superfield dependence of $A_a(x,\theta)$ leaving aside 
the first component $A_a(x,0)$. This happens because $ {\mathbb Y}^{(0|2)}$ as defined in (\ref{PCOb}) has an obvious non-trivial kernel.

 However, we can modify the PCO  
given in (\ref{PCOb}) with a more general construction. If we consider a set of anticommuting superfields $\Sigma^\a(x,\theta)$ 
such that $\Sigma^\a(x,0) = 0$. 
They can be normalised as $\Sigma^\a(x,\theta) = \theta^\a + K^\a(x, \theta)$ with $K^\a \approx {\cal O}(\theta^2)$. Then,we define 
\begin{equation}\label{PCOf}
{\mathbb Y}^{(0|2)} = \prod_{i=1}^m \Sigma^{\a_i} \delta(d \Sigma^{\a_i}) = 
\prod_{i=1}^m \Sigma^{\a_i} \delta\Big( (\delta^{\a_i}_\b + D_\b \Sigma^{\a_i}) d\theta^{\b} + V^a \partial_a\Sigma^{\a_i}\Big)
\end{equation}\,, 
$$
= \prod_{i=1}^m \Sigma^{\a_i}  
\delta\left[ (\delta^{\a_i}_\b + D_\b \Sigma^{\a_i}) \Big( d\theta^{\b} + V^a \frac{\partial_a\Sigma^{\b}}{(1 + D \Sigma)} \Big)\right]\,, 
$$
where $(1 + D\Sigma)$ is a $m \times m$ invertible matrix and it should be obvious from the above formula how the indices are contracted. 
Expanding the Dirac delta function and recalling that the bosonic dimension of the space is 3, we get the formula 
\begin{eqnarray}\label{PCOg}
{\mathbb Y}^{(0|2)} &=& H(x, \theta) \delta^2(d\theta) + K_a^{\a}(x,\theta) V^a \iota_\a \delta^2(d\theta) + \nonumber\\
&+& L_{ab}^{(\a\b)}(x,\theta) V^a V^b \iota_\a \iota_\b \delta^2(d\theta) + 
M_{abc}^{(\a\b\gamma)}(x,\theta) V^a V^b V^c \iota_\a \iota_\b \iota_\gamma \delta^2(d\theta)\,, 
\end{eqnarray}
where the superfields $H, K_a^{\a}, L_{ab}^{(\a\b)}$ and $M_{abc}^{(\a\b\gamma)}$ are easily computed in terms of $\Sigma^\a$ and its 
derivatives. Even if it is not obvious from the above expression, ${\mathbb Y}^{(0|2)}$ is closed and not exact. It belongs to $H^{(0|2)}$ 
and it is globally defined; this can be checked by decomposing the supermanifold in patches and checking that 
${\mathbb Y}^{(0|2)}$ is 
an element of the \v{C}ech cohomology, as carefully done in \cite{Catenacci:2010cs}. 
Now, if we compute the new field strength $\widetilde F^{(2|2)}$ by (\ref{PCOd}), one sees that the different pieces in 
(\ref{PCOg}) from ${\mathbb Y}$ are going to pick up different contributions from $F^{(2|0)}$. For instance, the $d\theta^\a \wedge d\theta^\b$ 
is soaked up from the third piece in (\ref{PCOg}) with the two derivatives acting on Dirac delta function.  

The choice of $\mathbb{Y}^{(0|2)}$ is the key of the present work, since the arbitrariness of the choice of 
$\mathbb{Y}^{(0|2)}$ allows us to relate the component action with the superspace formulation. 
For example, the new 
\begin{eqnarray}
\label{PCOga}
\mathbb{Y}^{(0|2)}_{new} = V^a \wedge V^b (\gamma_{ab})^{\a\b} \iota_\alpha \iota_\b \delta^2(\psi)\,. 
\end{eqnarray}
is closed as can be easily verified  
\begin{eqnarray}
\label{PCOgb}
d \mathbb{Y}^{(0|2)}_{new} = 2 \psi \gamma^a \psi V^b  (\gamma_{ab})^{\a\b} \iota_\alpha \iota_\b \delta^2(\psi) 
= {\rm tr}( \gamma^a \gamma_{ab}) V^b \delta^2(\psi) =0 \,. 
\end{eqnarray}
by using $d V^a = \psi \gamma^a \psi$ and $d\psi^\a =0$. 
It is not exact, it is invariant under rigid supersymmetry and it differs from ${\mathbb Y}^{(0|2)}$ by exact 
terms. This PCO can be expanded in different pieces by decomposing $V^a$ and by taking the derivatives $\iota_\a$ from 
$\delta^2(\psi)$ to $V$'s: 
\begin{eqnarray}
\label{SM-H}
\mathbb{Y}_{new}^{(0|2)}  = a_1 dx^a \wedge dx^b (\gamma_{ab})^{\a\b} \iota_\a \iota_\b \delta^2(\psi) + a_2 
 dx^a \wedge (\gamma^a \theta)^\b \iota_\b \delta^2(\psi) +
 a_3 \theta^2 \delta^2(\psi)
\,,
\end{eqnarray}
where the coefficients $a_i$ are fixed by simple Dirac matrix algebra. We notice that all pieces have 
zero form degree and picture number +2. Another property of ${\mathbb Y}^{(0|2)}_{new}$ is its duality with 
$\omega^{(3|0)} = \psi \gamma_a \psi V^a$. The latter is an element of the Chevalley-Eilenberg cohomology (see 
\cite{Castellani} for a complete discussion and references) and therefore it is closed (by using 
the Fierz identities $\gamma^a \psi (\psi\gamma_a \psi) =0$) and is not exact. The duality with ${\mathbb Y}^{(0|2)}_{new}$ 
means
\begin{eqnarray}
\label{SM-HA}
\omega^{(3|0)} \wedge {\mathbb Y}^{(0|2)}_{new} = \epsilon_{abc} V^a \wedge V^b \wedge V^c\,  \delta^2(\psi)\,, 
\end{eqnarray}
where $\epsilon_{abc} V^a \wedge V^b \wedge V^c \delta^2(\psi)$ is the volume form belonging to $\Omega^{(3|2)}$.

If the gauge group is non-abelian, the field strength $F^{(2|0)}$ has to be modified in 
\begin{equation}\label{PCOdA}
F^{(2|0)} = d A^{(1|0)} +  A^{(1|0)} \wedge A^{(1|0)} \,, 
\end{equation}
where the wedge product of two superform (at picture zero) gives a superform again at 
picture zero. 
However, to define a field strength at picture number 2, we immediately see 
that the product of $A^{(1|2)} \wedge A^{(1|2)} =0$ independently of the non-abelianity of the gauge group, 
but because $\delta^3(d\theta)=0\,.$

\section{Super Chern-Simons Action}
Let's begin by reviewing the standard superspace construction of Chern-Simons.
We start from a 1-super form $A^{(1|0)} = A_a V^a  + A_\a \psi^\a$, (where the superfields $A_a(x,\theta)$ and $A_\alpha(x,\theta)$ take value in the adjoint representation of the gauge group) and we define the field strength
\begin{eqnarray}\label{SM-A}
F^{(2|0)} = d A^{(1|0)} +  A^{(1|0)} \wedge A^{(1|0)} = F_{ab} V^a \wedge V^b + F_{a\alpha} V^a \wedge \psi^\a + F_{\a\b} \psi^\a\wedge \psi^\b \,,
\end{eqnarray}
where 
\begin{eqnarray}
\label{SM-B}
F_{ab} &=& \partial_a A_b - \partial_b A_a + [A_a, A_b]\,, ~~ \nonumber \\
F_{a\a} &=& \partial_a A_\alpha - D_\alpha A_a + [A_\alpha, A_b]\,, ~~ \nonumber \\
F_{\a\b} &=& D_{(\a} A_{\b)} + \gamma^a_{\a\b} A_a + \{A_\alpha, A_\beta\}\,, 
\end{eqnarray}
In order to reduce the redundancy of degrees of freedom because of the two 
components $A_a$ and $A_\a$ of the $(1|0)$ connection, one imposes (by hand) the conventional constraint 
\begin{eqnarray}
\label{SM-C}
\iota_\a \iota_\b F^{(2|0)} =0\,   ~~~ \Longleftrightarrow ~~~ F_{\a\b} = \nabla_{(\a} A_{\b)} + \gamma^a_{\a\b} A_a =0\,, 
\end{eqnarray}
from which it follows that $F_{a\alpha} = \gamma_{a, \alpha\beta} W^\beta$ with 
$W^\a = \nabla^\b \nabla^\a A_\b$ and $\nabla_\a W^\a =0$. 
The gaugino field strength $W^\a$ is gauge invariant 
under the non-abelian transformations $\delta A_\a = \nabla_\a \Lambda$. 
These gauge transformations descend from the gauge transformations $\delta A = \nabla \Lambda$ where $\Lambda$ is 
a $(0|0)$-form. 

The field strengths satisfy the following Bianchi's identities  
\begin{eqnarray}\label{SM-V}
&&\nabla_{[a} F_{bc]} =0 \,, \nonumber \\
&&\nabla_\a F_{ab} + (\gamma_{[a} \nabla_{b]} W)_\a =0\,, \nonumber\\
&&F_{ab} + \frac12 (\gamma_{ab})^\a_{~\beta} \nabla_\a W^\beta =0 \,, \nonumber \\
&&\nabla_\a W^\a =0\,. 
\end{eqnarray}
\newcommand{\sint}{\int\!\!\!\!\!\!-}
and by expanding the superfields $A_a, A_\a$ and $W^\a$ at the first components we 
have 
\begin{eqnarray}
\label{SM-VA}
A_\a = (\gamma^a \theta)_\alpha a_a + \lambda_\a \frac{\theta^2}{2}\,, ~~~~~~~
A_a = a_a + \lambda \gamma_a \theta + \dots \,, ~~~~~~
W^\a = \lambda^\a + f^\a_{~\b} \theta^\b + \dots\,, 
\end{eqnarray}
where $a_a(x)$ is the gauge field, $\lambda_\a(x)$ is the gaugino and $f_{\a\b} = \gamma_{\a\b}^{ab} f_{ab}$ 
is the gauge field strength with $f_{ab} = \partial_a a_b - \partial_b a_a$.  

In terms of those fields, the super-Chern-Simons lagrangian becomes 
\begin{eqnarray}
\label{SM-D}
S_{SCS} &=& \int {\rm Tr}  A_\a \left(W^\a - \frac{1}{6} [A_\b,A^a ]\gamma_a^{\a\b}\right) [d^3x d^2\theta]\,, 
\end{eqnarray}
which in component reads
\begin{eqnarray}
\label{SM-DA}
S_{SCS}  = \int d^3x {\rm Tr} \Big( \epsilon^{abc} (a_a \partial_b a_c + 
 \frac23 a_a a_b a_c) + \lambda_{\a} \epsilon^{\a\b} \lambda_\b\Big)\,. 
\end{eqnarray}
That coincides with  the bosonic Chern-Simons action  with free non-propagating fermions.  

In order to obtain an action principle by integration on supermanifolds we consider the natural candidates for 
a super-Chern-Simons lagrangian 
\begin{eqnarray}
\label{SM-DB}
{\cal L}^{(3|0)} = {\rm Tr} \left(A^{(1|0)} \wedge d A^{(1|0)} + \frac{2}{3}  A^{(1|0)}\wedge A^{(1|0)} \wedge A^{(1|0)} \right)\,, 
\end{eqnarray}
where $A^{(1|0)}$ is the superconnection and $d$ is the differential on the superspace, and then we multiply it 
by a PCO, for example by $\mathbb{Y}^{(0|2)} = \theta^2 \delta^2(d\theta)$. That leads to $(3|2)$ integral form that 
can be integrated on the supermanifold, that is 
\begin{eqnarray}
\label{SM-E}
S_{SCS} = 
\int_{SM} \mathbb{Y}^{(0|2)}\wedge {\rm Tr} \left(A^{(1|0)} \wedge d A^{(1|0)} + \frac{2}{3}  A^{(1|0)}\wedge A^{(1|0)} \wedge A^{(1|0)} \right)\,.
\end{eqnarray}
However, this action fails to give the correct answer yielding only the bosonic part of the action of $S_{SCS}$. 
The reason is that the supersymmetry transformations of the PCO is 
\begin{eqnarray}
\label{SM-F}
\delta_\epsilon \mathbb{Y} = d \left[ \theta^2 \iota_{\epsilon} \delta^2 (d\theta) \right]\,, 
\end{eqnarray}
and by integrating by parts, we find that the action is not supersymmetric invariant. 
On the other hand, as we observed in the previous section,  
we can use the new operator
\begin{eqnarray}
\label{SM-G}
\mathbb{Y}_{new}^{(0|2)} = V^a \wedge V^b (\gamma_{ab})^{\a\b} \iota_\a \iota_\b \delta^2(\psi)\,,
\end{eqnarray}
which is manifestly supersymmetric. Computing the expression in the integral, we see that 
${\mathbb Y}^{(0|2)}_{new}$ picks up al least two powers of $\psi$'s and one power of $V^a$ and that forces us 
to expand ${\cal L}^{(3|0)}$ as 3-form selecting the monomial $\psi \gamma_a \psi V^a$ dual to 
${\mathbb Y}^{(0|2)}_{new}$. 
Explicitly we find
\begin{eqnarray}\label{SM-GA}
S_{SCS} = \int {\rm Tr}\left(  A_\a F_{b\gamma}\gamma_{ac}^{\a\gamma}\epsilon^{abc}+A_a F_{\b\gamma}\gamma_{bc}^{\b\gamma}\epsilon^{abc} - \frac{1}{6}A_\a [A_\b,A^a ]\gamma_a^{\a\b}\right) [d^3x d^2\theta].
\end{eqnarray}
That finally gives the supersymmetric action described in (\ref{SM-D}), together with the conventional constraint $F_{\a\b}=0$.

Some observations are in order.
\begin{enumerate}
\item The equations of motion derived from the new action (\ref{introC}) 
are 
\begin{eqnarray}
\label{SM-IA}
&&\mathbb{Y}_{new}^{(0|2)}  (d A^{(1|0)} + A^{(1|0)} \wedge A^{(1|0)}) =0 ~~~~\Longrightarrow ~~~~ \nonumber \\
&&~~~~~~~~~V^3 (\gamma^a \iota)^\a \delta^2(\psi) F_{a\a} + 
(V^a \wedge V^b) \epsilon_{abc} (\gamma^c)^{\a\b} F_{\a\b} =0\,. 
\end{eqnarray}
The equations of motion correctly imply $F_{\a\b} =0$ (which is the conventional constraint) and $W^\a =0$ 
which are the super-Chern-Simons equations of motion. The second condition follows from $F_{\a\b} =0$  and by the Bianchi identities which implies that $F_{a\a} = \gamma_{a\a\b} W^\b$. 

Notice that this formulation allows us to get the conventional constraint as an equation of motion. In particular we find that
the equation of motion, together with the Bianchi identity 
imply the vanishing of the full field-strenght. 
\begin{eqnarray}
\left\{
\begin{array}{l}
\mathbb{Y}_{new}^{(0|2)} F^{(2|0)}=0,   \\
  ~   \\
 dF^{(2|0)}+[A^{(1|0)},F^{(2|0)}]=0,     
\end{array}
\right.  \quad\quad  \Longrightarrow \quad\quad F^{(2|0)}=0\,. 
\end{eqnarray}

\item Consider instead of the flat superspace $R^{(3|2)}$, the group manifold 
with the underlying supergroup ${\rm Osp}(1|2)$.  The corresponding Maurer-Cartan equations are 
\begin{eqnarray}
\label{SM-L}
d V^a + \epsilon^{a}_{~bc} V^b\wedge V^c +  \psi \gamma^a \psi=0\,, ~~~~~
d \psi^\a + (\epsilon\gamma_a)^{\a}_{~\b} V^a \psi^\b =0\,.  
\end{eqnarray}
Then, it is easy to show that 
\begin{eqnarray}
\label{SM-M}
d \mathbb{Y}_{new}^{(0|2)} = 0\,, ~~~~
\delta_\epsilon \mathbb{Y}_{new}^{(0|2)} =0\,.
\end{eqnarray}
The second equation is obvious since it is expressed in terms of supersymmetric invariant quantities. 
The first equation follows from the MC equations and gamma matrix algebra. Chern-Simons theory on this group 
supermanifold share interesting similarities with a particular version of open super string field theory \cite{PTY}. The reason for this is that the supergroup Osp$(1|2)$ is infact the superconformal Killing group of an $N=1$ SCFT on the disk. There 
is however an important difference wrt to \cite{PTY}. Our choice of the picture changing operator $\mathbb{Y}$ 
applied to the field strength 
$ (d A^{(1|0)} + A^{(1|0)} \wedge A^{(1|0)})$ leads to equation (\ref{SM-IA}) and it directly implies the vanishing of the 
full field strength. In particular the kernel of the picture-changing operator is harmless in our case. It would be interesting to search 
for an analogous object in the RNS string.

\item The PCO $\mathbb{Y}_{new}^{(0|2)}$ is related to the product of two non-covariant operators, each shifting the picture by one unit.
\begin{eqnarray}
\label{SM-N}
Y_v = V^a v_\a \gamma^{\a\b}_a \iota_\b  \delta( v \cdot \psi)\,, ~~~~~~~
Y_w = V^a w_\a \gamma^{\a\b}_a \iota_\b  \delta( w \cdot \psi)\,, ~~~~
 \end{eqnarray}
with $ v\cdot w \neq 0$ and by a little a bit of algebra, one gets 
\begin{eqnarray}
\label{SM-NA}
\mathbb{Y}_{new}^{(0|2)}  = Y_v  Y_w  + d \Omega\,.
\end{eqnarray}
 The PCO's $Y_v$ and $Y_w$ are closed (in the case of 
flat superspace, while in the case of ${\rm Osp}(1|2)$, they are invariant if $v$ and $w$ transform under 
the corresponding isometry transformations). They are also supersymmetric invariant because written in terms 
of invariant quantities. 

The piece $\Omega$ is a $(-1|2)$ form which depends on $v$ and $w$. The two PCO's are equivalent in 
the sense that they belong to  the same cohomology class, but they behave differently off-shell. One can check 
by direct inspection that this PCO does not lead to the conventional constraint $F_{\a\b} =0$ and therefore 
the exact term in (\ref{SM-NA}) relating the two actions is important to get the full-fledged action principle.  

\item We study the kernel of the PCO $\mathbb{Y}^{(0|2)}$ and of the new PCO 
$\mathbb{Y}^{(0|2)}_{new}$. 

Acting on the complete set of differential form $\Omega^{(p|q)}$, 
with the PCO's,  for $\omega^{(p|q)} \in \Omega^{(p|q)}$ with $q>0$, we have 
$\mathbb{Y}^{(0|2)} \wedge \omega^{(p|q)} =0$ due to the anticommuting properties of $\delta(d\theta)$. 
Therefore, we need to study only $\Omega^{(p|0)}$. 
We observe that $\mathbb{Y}^{(0|2)} \wedge \omega^{(0|0)} =0$, this implies 
$\omega^{(0|0)} = f_{1,\alpha}(x) \theta^\alpha + f_{2}(x) \theta^2$. In the same way, given a $1$-form of 
$\Omega^{(1|0)}$, we have $\omega^{(1|0)} = \omega_a(x,\theta) V^a + \omega_{\a}(x, \theta) d\theta^\alpha$. 
Then, the kernel of $\mathbb{Y}^{(0|2)}$  on $\Omega^{(1|0)}$ is 
given by 
\begin{eqnarray}
\label{kerA}
\omega^{(1|0)} =  \Big(\omega_{1,a \alpha}(x) \theta^\alpha  + \omega_{2, a}(x) \theta^2\Big) V^a + 
\omega_{\a}(x, \theta) d\theta^\alpha\,. 
\end{eqnarray}
For higher $p$-forms, we have similar kernels. For instance, in the case of $2$-forms $\Omega^{(2|0)}$, 
we have 
\begin{eqnarray}
\label{kerB}
\omega^{(2|0)} &=& \Big(\omega_{1, a b \a}(x) \theta^\a + \omega_{2, a b}(x) \theta^2\Big) V^a \wedge V^b 
\nonumber \\ 
&+& 
\omega_{a \alpha}(x, \theta) V^a \wedge d\theta^\alpha  + 
\omega_{\alpha \beta}(x, \theta) d\theta^\a \wedge d\theta^\b\,, 
 \end{eqnarray}

Let us study the kernel of the new PCO $\mathbb{Y}^{(0|2)}_{new}$. On $\Omega^{(0|0)}$, there is 
no kernel. Acting on 
$\omega^{(1|0)} =  \omega_a(x,\theta) V^a + \omega_{\a}(x, \theta) d\theta^\alpha$, we have 
\begin{eqnarray}
\label{kerC}
\mathbb{Y}^{(0|2)}_{new} \wedge \omega^{(1|0)} &=& V^3 \epsilon^{abc} \omega_{c}(x,\theta) 
(\gamma_{ab})^{\a\b} \iota_\alpha \iota_\b \delta^2(d\theta) \nonumber \\ 
&+& 
2 V^a\wedge V^b   (\gamma_{ab})^{\a\b} \omega_\alpha(x,\theta) \iota_\b \delta^2(d\theta) =0\,.
\end{eqnarray}
Since the two forms $V^a\wedge V^b (\gamma_{ab})^{\a\b} \iota_\b \delta^2(d\theta)$ and 
$ V^3 \epsilon^{abc} (\gamma_{ab})^{\a\b} \iota_\alpha \iota_\b \delta^2(d\theta)$ generate 
the space $\Omega^{(1|2)}$ (which has dimension $(3|2)$), the kernel of  $\mathbb{Y}^{(0|2)}_{new}$ is 
given by the solution of 
\begin{eqnarray}
\label{kerD}
\epsilon^{abc} \omega_{c}(x,\theta) 
(\gamma_{ab})^{\a\b} =0\,, ~~~~~~
\epsilon^{abc} (\gamma_{ab})^{\a\b} \omega_\alpha(x,\theta) =0 
\end{eqnarray}
which imply that $\omega_c(x,\theta) = \omega_\alpha(x, \theta) =0$. Thus, there is no kernel on 
$\Omega^{(1|0)}$.  We move to the more important class: $\Omega^{(2|0)}$. 
For that we consider the generic $2$-form, and the kernel equation gives 
\begin{eqnarray}
\label{kerF}
\gamma_{ab}^{\a\b} \omega_{\a\b}(x,\theta) = 0\,, ~~~~~~
\gamma_{ab}^{\a\b} \epsilon^{abc} \omega_{c \alpha}(x,\theta) = 0\,. 
\end{eqnarray}
No condition imposed on $\omega_{ab}(x,\theta)$. The first equation implies that $\omega_{\a\b}(x,\theta)=0$, 
while, by decomposing  
$\omega_{c \alpha}(x,\theta) = (\gamma_{c})^{\b\gamma} \widetilde\omega_{\a\b\gamma} + 
(\gamma_{c})_{\a\b} \hat\omega^\b$ where $\widetilde\omega_{\a\b\gamma}(x,\theta)$ is totally symmetric in 
the spinorial indices, we have $ \hat\omega^\b =0$. The reason why ${\mathbb Y}^{(0|2)}_{new}$ works in the construction of an action is that the 
$\widetilde\omega_{\a\b\gamma}(x,\theta)$ component of the field strength is independently set to zero by the Bianchi identity. In the same way, one can analyze further higher $p$-forms. 


\end{enumerate}

\section{Changing the PCO and the relation between different superspace formulations}

During the last thirty years, we have seen two independent superspace formalisms taking place, 
aiming to describe supersymmetric theories from a geometrical point of view. They are known as as {\it 
superspace} technology, whose basic ingredients are collected in series of books (see for example 
\cite{Gates:1983nr,Wess:1992cp}) and the {\it rheonomic} 
(also known as {\it group manifold}) formalisms 
(see the main reference book \cite{Castellani}). They are based on a  different approach and they 
have their own advantages and drawbacks. Without entering the details of those formalisms, we would like to illustrate 
some of their main  features on the present example of super-Chern-Simons theories. A basic difference is that in the superspace few superfields contain the basic fields of the theory as components, while in the rheonomic approach any basic field of the theory is promoted to a superfield. 

Let us start from the rheonomic action. 
This is given as follows
\begin{eqnarray}
\label{reA}
S_{rheo}[A, {\cal M}^3] = \int_{{\cal M}^3 \subset {\cal SM}^{(3|2)}} {\cal L}^{(3)}(A)\,, 
\end{eqnarray}
where ${\cal M}^3$ is a three-dimensional surface immersed into the supermanifold 
${\cal SM}^{(3|2)}$ and ${\cal L}^{(3)}(A)$ is a three-form Lagrangian constructed with superform $A$, 
their derivatives without the Hodge dual operator (that is without any reference to a metric on the 
supermanifold ${\cal SM}^{(3|2)}$). Notice that the fields $A$ are indeed superforms whose 
components are superfiels. 

The action $S_{rheo}[A, {\cal M}^3]$ 
is a functional of the superfields and of the embedding of ${\cal M}^3$ into ${\cal SM}^{(3|2)}$. We can 
then consider the classical equations of motion by minimizing the action both respect to the variation 
of the fields and of the embedding. However, the variation of the immersion can be compensated by  
diffeomorhisms on the fields if the action ${\cal L}^{(3)}$ is a differential form. This implies that the complete set of 
equations associated to the action (\ref{reA}) are the usual equations obtained by varying the fields on a fixed surface 
${\cal M}^{3}$ with the proviso that these equations hold not only on ${\cal M}^3$, but on the whole supermanifold ${\cal M}^{(3|2)}$, namely the Lagrangian is a function of $(x^a, \theta^\alpha, V^a, \psi^\alpha)$. 

The rules to build the action (\ref{reA}) are listed and discussed in the book \cite{Castellani} in  detail. 
An important ingredient is the fact that for the action to be supersymmetric invariant,  
the Lagrangian must be invariant up to a $d$-exact term and, in addition, if the algebra of supersymmetry closes off-shell 
(either because there is no need of auxiliary fields or because it exists a formulation with auxiliary fields), the 
Lagrangian must be closed: $d {\cal L}^{(3)}(A) =0$, {\it upon using the rheonomic parametrization}. One of the
rules of the geometrical construction for supersymmetric theories given in \cite{Castellani} is 
that by setting to zero the coordinates $\theta^\a$ and its differential $\psi^\a = d\theta^\a$, the action 
\begin{eqnarray}
\label{reB}
S_{rheo}[A] = \int_{{\cal M}^3}  \left. {\cal L}^{(3)}(A) \right|_{\theta =0, d\theta =0}\,,
\end{eqnarray}
 reduces to the component action invariant under supersymmetry. Furthermore, 
 the equations of motion in the full-fledged superspace implies the rheonomic constraints (which coincide with the conventional constraints of superspace formalism). 

In order to express the action (\ref{reA}) in a more geometrical way by including the dependence upon the 
embedding into the integrand, we refer to \cite{Castellani:2015paa} and we introduce the Poincar\'e dual form 
$\mathbb{Y}^{(0|2)}= \theta^2 \delta^2(d\theta)$. As already discussed in the previous section, 
$\mathbb{Y}^{(0|2)}$ is closed and 
its supersymmetry variation is $d$-exact. The action can be written on the full supermanifold as
\begin{eqnarray}
\label{reC}
S[A] = \int_{{\cal SM}^{(3|2)} } {\cal L}^{(3|0)}(A)\wedge  \mathbb{Y}^{(0|2)}\,,
\end{eqnarray}
Therefore, by choosing the PCO ${\mathbb Y}^{(0|2)} = \theta^2 \delta^2(d\theta)$, 
its factor $\theta^2$ projects the Lagrangian $ {\cal L}^{(3|0)}(A)$ to ${\cal L}^{(3)}(A)_{\theta=0}$
while the factor $\delta^2(d\theta)$ projects the latter to  ${\cal L}^{(3)}(A)_{\theta =0, d\theta=0}$ reducing 
$S[A]$ to the component action (\ref{SM-DA}).  

Any variation of the embedding yields $\delta \mathbb{Y}^{(0|2)} = d \Lambda^{(-1|2)}$ leaves the action invariant if the 
Lagragian is closed. In the case of Chern-Simons discussed until now, the chosen action was identified only with 
the bosonic term $A\wedge dA$, but that turns out to be not closed. Therefore, that has to be modified as follows: 
as discussed above the physical fields of Chern-Simons theory are the gauge field $a_\mu$ and the gaugino 
$\lambda_\alpha$ 
which are the zero-order components of the supergauge field $A(x,\theta)$ and of the spinorial superfield 
$W^\a(x, \theta)$, the complete closed action reads
\begin{eqnarray}
\label{reD}
S[A] = \int_{{\cal SM}^{(3|2)}} \Big( A\wedge dA + \frac23 A\wedge A\wedge A + W^\a W_\a V^3  \Big)\wedge {\mathbb Y}^{(0|2)}\,, 
\end{eqnarray}
which is a $(3|2)$ form.\footnote{This $(3|0)$ Lagrangian in (\ref{reD}) already appeared in \cite{Fabbri:1999ay} by reducing their formula from $N=2$ to $N=1$.} 
Imposing the closure of ${\cal L}^{(3|0)}$ we get the rheonomic parametrizations of the curvatures, or differently said, the conventional constraints. Once this is achieved, we are free to choose  any PCO in the same cohomology class.
If we choose  the PCO ${\mathbb Y}^{(0|2)} = \theta^2 \delta^2(d\theta)$ we get directly the 
component action (\ref{SM-DA}) and the third term in the action is fundamental to get the mass term for the non-dynamical fermions. On the other hand, by choosing ${\mathbb Y}^{(0|2)}_{new}$, (\ref{introB}) we see that the last term is unessential becasue, due to the powers 
of $V^a$, this term cancels out and we get the superspace action (\ref{SM-D}). 

This is the most general action and the closure of ${\cal L}^{(3|0)}$ implies that any gauge invariant 
and supersymmetric action can be built by choosing $\mathbb{Y}^{(0|2)}$ inside of the same cohomology class. Therefore, 
starting from the rheonomic action, one can choose a different ``gauge" -- or better said a different embedding of the 
submanifold ${\cal M}^3$ inside the supermanifold ${\cal SM}^{(3|2)}$ -- leading to different forms of the action with the same physical content. 
It should be stressed, however, that the choice of ${\mathbb Y}^{(0|2)}_{new}$, (\ref{introB}), is a preferred ``gauge'' choice, which allows us to derive the conventional constraint  by varying the action without using the rheonomic parametrization.

\vspace{ 1cm} 
\noindent\textbf{Acknowledgements}

\noindent We thank L. Castellani, R. Catenacci,  T. Erler and G. Policastro 
for valuable discussions. 
The research of CM is funded by a {\it Rita Levi Montalcini} grant from the Italian MIUR.


\end{document}